\documentclass[11pt]{article}

\usepackage{graphicx} 
\usepackage{amsmath}
\usepackage{amsfonts}
\usepackage{amssymb}

\begin{document}

\title {$\bar{D}$-brane as Dark Matter in Warped String Compactification}
\author{Shinji Mukohyama\\
\\
{\it Department of Physics}\\
{\it and}\\
{\it Research Center for the Early Universe}\\
{\it The University of Tokyo, Tokyo 113-0033, Japan}\\
}
\date{\today}

\maketitle

\abstract{
It is pointed out that in the warped string compactification, motion of
$\bar{D}$-branes near the bottom of a throat behaves like dark
matter. Several scenarios for production of the dark matter are
suggested, including one based on the $D/\bar{D}$ interaction at the
late stage of $D/\bar{D}$ inflation. 
\begin{flushright}
 UTAP-525, RESCEU-8/05
\end{flushright}
}

\section{Introduction}

M/string theory~\cite{GSW,Polchinski} is considered as a strong
candidate for a unified theory of fundamental physics. Its mathematical
consistency and beauty have been attracting interest of many
physicists. On the other hand, one of the drawbacks is lack of direct
experimental or observational evidence of such a structure at high
energies. Having this situation, it seems rather natural to turn our
eyes to cosmology and look for cosmological implication of M/string
theory since the universe is supposed to have experienced a high energy 
epoch at its early stage.

Considering the success of inflation~\cite{Sato,Guth,Linde} as a
scenario of the early universe and the observational evidences for
accelerating expansion of the present
universe~\cite{Perlmutter,Schmidt,Riess}, one of the important steps 
toward establishment of M/string cosmology would be construction of de
Sitter or quasi-de Sitter universe. However, for a long time it seemed
rather difficult to construct $4$-dimensional de Sitter universe in
M/string theory, especially if we seriously take account of the moduli
stabilization. Indeed, the no-go theorem of \cite{WSD,Maldacena-Nunez}
says that in a large class of supergravity theories, there is no
no-singular (warped) compactification to $4$-dimensional de Sitter space
with a finite $4$-dimensional Newton's constant.

There are several proposals to evade the no-go theorem by inclusion of
additional sources such as stringy corrections to the supergravity
theories in the $g_s$ or $\alpha'$ expansion and extended sources,
i.e. branes. The recent proposal by Kachru, Kallosh, Linde and
Trivedi (KKLT)~\cite{KKLT} utilizes various ingredients of string theory
including warped geometry, fluxes, $D$-branes, $\bar{D}$-branes,
instanton corrections to moduli potential in order to construct a
meta-stable de Sitter vacua in string theory~\footnote{
See \cite{Silverstein} and \cite{Townsend-Wohlfarth,Ohta} for other
proposals of de Sitter and transiently accelerating universe.
}.

In the KKLT setup $\bar{D}3$-branes play an essential role. Inclusion of
$\bar{D}3$-branes at the bottom of a warped throat uplifts stable AdS
vacua with negative cosmological constant to meta-stable de Sitter vacua
with positive cosmological constant in a theoretically controllable
way. Without $\bar{D}3$-branes, we would end up with a negative
cosmological constant, which is inconsistent with observations.

The purpose of this paper is to point out that motion of the 
$\bar{D}3$-branes near the bottom of a warped throat behaves like dark
matter and that it can be naturally generated in the context of brane
cosmology. This paper is organized as follows. In Sec.~\ref{sec:KS} we
briefly review the Klebanov-Strassler geometry which approximates the
geometry in a warped throat, and consider a probe $\bar{D}3$-brane. In 
Sec.~\ref{sec:cosmology} we promote the $\bar{D}3$-brane action
formulated in a $4$-dimensional flat background to a curved background
and investigate its implications to cosmology. In particular, we show
that a $\bar{D}3$-brane near the bottom of a warped throat behaves like
dark matter. In Sec.~\ref{sec:production} we suggest several possible
scenarios to produce the dark matter. Finally, Sec.~\ref{sec:summary} is
devoted to a summary of this paper and discussion.

\section{$\bar{D}3$-brane in Klebanov-Strassler geometry} 
\label{sec:KS}

In the KKLT setup~\cite{KKLT} a throat region is described by the warped
deformed conifold solution of Klebanov and
Strassler~\cite{Klebanov-Strassler}. To be precise, the warped deformed
conifold geometry is compactified by additional fluxes as described by 
Giddings, Kachru and Polchinski~\cite{GKP}. Hence, the geometry in the
UV region of the throat significantly deviates from the
Klebanov-Strassler solution, but the geometry near the bottom of the
throat, i.e. in the IR region, is approximated by the Klebanov-Strassler
solution.

The Klebanov-Strassler geometry~\cite{Klebanov-Strassler,CHKO} has the
simple ansatz: 
%
\begin{equation}
 ds^2 = h^{-1/2}(\tau)\eta_{\mu\nu}dx^{\mu}dx^{\nu}+ h^{1/2}(\tau)ds_6^2,
\end{equation}
where $x^{\mu}$ ($\mu=0,\cdots,3$) are $4$-dimensional coordinates and 
$ds_6^2$ is the metric of the deformed
conifold~\cite{Candelas-Ossa,Minasian-Tsimpis} 
%
\begin{equation}
 ds_6^2 = \frac{\epsilon^{4/3}}{2}K(\tau)
  \left[ \frac{1}{3K^3(\tau)}\left(d\tau^2+(g^5)^2\right)
   + \cosh^2\left(\frac{\tau}{2}\right)\left((g^3)^2+(g^4)^2\right)
   + \sinh^2\left(\frac{\tau}{2}\right)\left((g^1)^2+(g^2)^2\right)
  \right]. \label{eqn:metric-deformed-conifold}
\end{equation}
Here, 
%
\begin{equation}
 K(\tau)= \frac{(\sinh(2\tau)-2\tau)^{1/3}}{2^{1/3}\sinh\tau},
\end{equation}
and $g^i$ ($i=1,\cdots,5$) are orthonormal basis defined by 
%
\begin{eqnarray}
 g^1 & = & \frac{e^1-e^3}{\sqrt 2},\quad
  g^2 = \frac{e^2-e^4}{\sqrt 2}, \nonumber \\
 g^3 & = & \frac{e^1+e^3}{\sqrt 2},\quad
  g^4 = \frac{e^2+ e^4}{\sqrt 2},\quad 
  g^5 = e^5,
\end{eqnarray}
where
%
\begin{eqnarray}
 e^1 & \equiv & - \sin\theta_1 d\phi_1, \quad
  e^2 \equiv d\theta_1, \nonumber \\
 e^3 & \equiv & 
  \cos\psi\sin\theta_2 d\phi_2-\sin\psi d\theta_2, \nonumber\\
 e^4 & \equiv & \sin\psi\sin\theta_2 d\phi_2+\cos\psi d\theta_2, 
  \nonumber \\
 e^5 & \equiv & d\psi + \cos\theta_1 d\phi_1+ \cos\theta_2 d\phi_2.
\end{eqnarray}
Because of the warp factor $h^{-1/2}(\tau)$, this geometry is often
called the warped deformed conifold. The R-R $3$-form field strength
$F_3$ and the NS-NS $2$-form potential $B_2$ also have the $Z_2$
symmetric (($\theta_1$, $\phi_1$) $\leftrightarrow$ ($\theta_2$,
$\phi_2$)) ansatz: 
%
\begin{eqnarray}
 F_3 & = & \frac{M\alpha'}{2} \left\{g^5\wedge g^3\wedge g^4 + d [ F(\tau)
	 (g^1\wedge g^3 + g^2\wedge g^4)]\right\}, \nonumber\\
 B_2 & = & \frac{g_s M\alpha'}{2}
  [f(\tau) g^1\wedge g^2 +  k(\tau) g^3\wedge g^4 ],
\end{eqnarray}
where $F(0)=0$ and $F(\infty)=1/2$. For this ansatz with the additional
condition 
%
\begin{equation}
 g_s^2F_3^2 = H_3^2,
\end{equation}
we can consistently set the dilaton $\phi$ and the R-R scalar $C_0$
to zero. The BPS saturated solution found by Klebanov and
Strassler~\cite{Klebanov-Strassler} is
%
\begin{eqnarray}
 F(\tau) & = & \frac{\sinh\tau -\tau}{2\sinh\tau}, \nonumber\\
 f(\tau) & = & \frac{\tau\coth\tau-1}{2\sinh\tau}(\cosh\tau-1), \nonumber\\
 k(\tau) & = & \frac{\tau\coth\tau-1}{2\sinh\tau}(\cosh\tau+1),
\end{eqnarray}
and 
%
\begin{equation}
 h(\tau) = 2^{2/3}\cdot (g_sM\alpha')^2\epsilon^{-8/3}I(\tau), 
\end{equation}
where 
%
\begin{equation}
 I(\tau) = 
  \int_\tau^\infty dx \frac{x\coth x-1}{\sinh^2 x}
  (\sinh(2x)-2x)^{1/3}.
\end{equation}
For this solution, 
%
\begin{equation}
 C_4 = h^{-1}dx^0\wedge dx^1\wedge dx^2\wedge dx^3
\end{equation}
in a particular gauge. For large $g_sM$ the curvature is small
everywhere and we can trust the supergravity description.

When $g_sM$ is sufficiently large, we can treat a $\bar{D}3$-brane as a
probe brane~\cite{Kachru-McAllister}. The action for the probe 
$\bar{D}3$-brane is 
%
\begin{equation}
 S_{\bar{D}3} = -T_3\int d^4\xi e^{-\phi}
  \sqrt{-\det(G_{\alpha\beta}-B_{\alpha\beta})}
  - T_3\int d^4\xi C_4,
\end{equation}
where $\xi^{\alpha}$ ($\alpha=0,\cdots,3$) are intrinsic coordinates on
the $\bar{D}3$-brane, $T_3$ is the tension and 
%
\begin{equation}
 G_{\alpha\beta} = G_{MN}
  \frac{\partial x^M}{\partial\xi^{\alpha}}
  \frac{\partial x^N}{\partial\xi^{\beta}}, \quad
 B_{\alpha\beta} = (B_{2})_{MN}
  \frac{\partial x^M}{\partial\xi^{\alpha}}
  \frac{\partial x^N}{\partial\xi^{\beta}}.
\end{equation}
In the following we shall adopt a gauge in which brane coordinates
$\xi^{\alpha}$ coincide with $x^{\alpha}$:
%
\begin{equation}
 x^{\alpha} = \xi^{\alpha},\quad \psi^m = \psi^m(\xi^{\alpha}), 
\end{equation}
where $\{\psi^m\}$ ($m=5,\cdots,10$) represents
$\{\tau,\psi,\theta_1,\phi_1,\theta_2,\phi_2\}$. In the
non-relativistic limit, 
%
\begin{equation}
 S_{\bar{D}3} = T_3\int d^4\xi 
  \left[-\frac{1}{2}\gamma_{mn}\eta^{\alpha\beta}
   \frac{\partial\psi^m}{\partial\xi^{\alpha}}
   \frac{\partial\psi^n}{\partial\xi^{\beta}}
   -2h^{-1} \right].
  \label{eqn:effective-action}
\end{equation}
where $\gamma_{mn}d\psi^md\psi^n=ds_6^2$ is the metric of the deformed
conifold given by (\ref{eqn:metric-deformed-conifold}).

\section{Cosmological implication}
\label{sec:cosmology}

We have considered a probe $\bar{D}3$-brane in the Klebanov-Strassler
geometry and have obtained the effective action
(\ref{eqn:effective-action}). Let us now consider cosmological
implication of this action~\footnote{
Another approach to cosmology with the warped string compactification
would be to consider time-dependent extension of the Klebanov-Strassler
solution. However, without compactification~\cite{GKP} and moduli
stabilization~\cite{KKLT}, there are instabilities associated with the
unfixed moduli~\cite{Kodama-Uzawa}.
}.
With moduli stabilization, the $4$-dimensional Einstein gravity should
be recovered~\footnote{
Evidence can be seen for the recovery of the $4$-dimensional Einstein
gravity in a simplified setup of warped flux
compactification~\cite{MSYK}.
}. 
In cosmological situations, we still expect that the KS geometry is a
good approximation to the geometry of extra dimensions in the throat
region as far as the energy scale associated with the $4$-dimensional
physics is sufficiently lower than the stabilization scale. Hence, we
promote the action (\ref{eqn:effective-action}) formulated in the flat
$4$-dimensional spacetime to a curved background
$ds_4^2=g^{(4)}_{\alpha\beta}d\xi^{\alpha}d\xi^{\beta}$ as  
%
\begin{equation}
 S_{\bar{D}3} = T_3\int \sqrt{-g^{(4)}}d^4\xi 
  \left[-\frac{1}{2}\gamma_{mn}g^{(4)\alpha\beta}
   \frac{\partial\psi^m}{\partial\xi^{\alpha}}
   \frac{\partial\psi^n}{\partial\xi^{\beta}}
   -2h^{-1} \right].
\end{equation}
We expect that this action describes the dynamics of the brane at energy
scales sufficiently lower than the stabilization scale. In general,
stabilization of the volume modulus introduces corrections to the
effective action, in particular to the mass of the fields. Actually, in 
Ref.~\cite{KKLMMT} it was pointed out that in the context of
$D3/\bar{D}3$ inflation, the inflaton mass receives corrections of order
$H$ due to the modulus stabilization. Those corrections were significant
for the inflaton dynamics since the inflaton mass must be fine-tuned to
a value much smaller than $H$ in order to realize a sufficient
inflation. On the other hand, our interest in this paper is on a massive
field (which shall be denoted by $\varphi$ below). As far as the mass of
the field of interest is much larger than the Hubble expansion rate $H$,
the corrections due to the modulus stabilization should be ignorable.

With the FRW ansatz 
%
\begin{equation}
 ds_4^2 = -dt^2 + a(t)^2d{\bf x}^2, \quad \psi^m=\psi^m(t),
\end{equation}
the equation of motion of $\psi^m$ is
%
\begin{equation}
 \dot{\pi}_m + 3H\pi_m + 
  \frac{\partial}{\partial\psi^m}\rho(\psi,\pi) = 0,
\end{equation}
where a dot denotes the time derivative, 
%
\begin{equation}
  \pi_m \equiv T_3\gamma_{mn}\dot{\psi}^n,
\end{equation}
and
%
\begin{equation}
 \rho(\psi,\pi) \equiv \frac{1}{2T_3}\gamma^{mn}\pi_m\pi_n
  + 2T_3 h^{-1} 
\end{equation}
is the energy density written in terms of $\psi^m$ and $\pi_m$.

Hereafter, we consider the $\bar{D}3$-brane near the bottom of the
throat $\tau=0$. Thus, we take the small $\tau$ limit of the deformed
conifold metric: 
%
\begin{equation}
 \gamma_{mn}d\psi^md\psi^n = 
  \frac{\epsilon^{4/3}}{2}\left(\frac{2}{3}\right)^{1/3}
  \left\{
  \frac{1}{2}d\tau^2 + 
  \left[\frac{1}{2}\left(g^5\right)^2
   + \left(g^3\right)^2 + \left(g^4\right)^2  
  \right]
  + \frac{1}{4}\tau^2
  \left[\left(g^1\right)^2 + \left(g^2\right)^2 \right]
	\right\},
\end{equation}
where the first square bracket represents the regular $S^3$ at the
bottom of the throat and the second square bracket represents the
shrinking $S^2$. With this form of the deformed conifold metric, it is
shown by using the equations of motion that there are constants of
motion $J_3^2$ and $J_2^2$ related to the angular momenta along the
$S^3$ and the $S^2$, respectively. For later convenience the constants of
motion $J_3^2$ and $J_2^2$ are normalized as 
%
\begin{eqnarray}
 \left(\frac{2}{\epsilon^{4/3}T_3}\right)^2
  \left(\frac{3}{2}\right)^{1/3}\frac{J_3^2}{2a^6} & \equiv & 
  \frac{1}{2}\left(g^5(\partial_t)\right)^2
   + \left(g^3(\partial_t)\right)^2 + \left(g^4(\partial_t)\right)^2,
   \nonumber\\
 \left(\frac{8}{\epsilon^{4/3}T_3}\right)^2
  \left(\frac{3}{2}\right)^{2/3}\frac{J_2^2}{2a^6\tau^4} & \equiv & 
 \left(g^1(\partial_t)\right)^2 + \left(g^2(\partial_t)\right)^2. 
\end{eqnarray}
It is shown that $a^3\pi_{\phi_1}$ and $a^3\pi_{\phi_2}$ are also
constants of motion, but they do not appear in the following
discussions. The equation of motion for $\tau$ and the energy density 
$\rho$ are written in terms of $J_3^2$ and $J_2^2$ as
%
\begin{eqnarray}
 \ddot{\varphi} + 3H\dot{\varphi} + 
  \left(\frac{\partial V}{\partial \varphi}\right)_a =0,
  \label{eqn:eom-varphi}\\
 \rho = \frac{1}{2}\dot{\varphi}^2 + V + \rho_6 + \rho_0,
\end{eqnarray}
where
%
\begin{eqnarray}
 \varphi & \equiv & \frac{\sqrt{\epsilon^{4/3}T_3}}{2}
  \left(\frac{2}{3}\right)^{1/6}\tau,\nonumber\\
 V(\varphi,a) & = & \frac{1}{2}m^2\varphi^2 
 + \frac{J_2^2}{2a^6\varphi^2}, \nonumber\\
 m^2 & = & \frac{8}{\epsilon^{4/3}}
  \left(\frac{3}{2}\right)^{1/3}
  \left.\frac{d^2}{d\tau^2}h^{-1}\right|_{\tau=0}
  = \frac{4\cdot 2^{2/3}}{3I(0)^2}\frac{\epsilon^{4/3}}{(g_sM\alpha')^2},
  \label{eqn:m2}
\end{eqnarray}
and
%
\begin{eqnarray}
 \rho_6 & = & \frac{1}{\epsilon^{4/3}T_3}\frac{J_3^2}{2a^6}, \nonumber\\
 \rho_0 & = & 2T_3\left.h^{-1}\right|_{\tau=0}.
\end{eqnarray}
Here, $I(0)\simeq 0.71805$.

For $J_2\ne 0$, the potential $V$ is minimized if $\varphi$ evolves as 
%
\begin{eqnarray}
 \varphi = \sqrt{\left|\frac{J_2}{m}\right|}\frac{1}{a^{3/2}}. 
  \label{eqn:DM-like-solution}
\end{eqnarray}
Actually, this is a solution to the equation of motion
(\ref{eqn:eom-varphi}) if the FRW universe evolves as 
$H\propto a^{-3/2}$. (We shall confirm this cosmological evolution
below.) With this solution, the energy density $\rho$ is 
%
\begin{equation}
 \rho = \left(1+\frac{9H^2}{8m^2}\right)\rho_3 + \rho_6 + \rho_0,
\end{equation}
where
%
\begin{equation}
 \rho_3 = \frac{|mJ_2|}{a^3}. 
\end{equation}
At low energy we can safely neglect $9H^2/8m^2$ compared to $1$. The
term $\rho_6$ can also be neglected since it decays faster than other
terms as the universe expands. Therefore, we obtain  
%
\begin{equation}
 \rho \simeq \rho_3 + \rho_0. 
\end{equation}
The first term behaves like dark matter while the second term
contributes to the cosmological constant. In this paper we shall not try
to solve the cosmological constant problem and simply assume that
$\rho_0$ is (almost) canceled by other contributions to the cosmological
constant (eg. stabilization of the volume modulus, tension of other
branes, etc.). With this assumption, the cosmological energy density is
dominated by $\rho_3$ and it is confirmed that the cosmological
evolution is $H\propto a^{3/2}$ as assumed, provided that the Friedmann
equation is recovered at low energy. Now let us consider a small
perturbation around the solution (\ref{eqn:DM-like-solution}) for
$J_2\ne 0$: 
%
\begin{eqnarray}
 \varphi = \sqrt{\left|\frac{J_2}{m}\right|}\frac{1}{a^{3/2}}
  + \delta\varphi(t).
\end{eqnarray}
The linearized equation of motion implies that the perturbation
$\delta\varphi$ behaves like a massive scalar field with mass $2m$, 
%
\begin{eqnarray}
 \delta\ddot{\varphi} + 3H\delta\dot{\varphi} + (2m)^2\delta\varphi = 0,
\end{eqnarray}
and that the solution (\ref{eqn:DM-like-solution}) is indeed stable. At
low energy $H\ll m$, the stress energy tensor due to $\delta\varphi$ 
behaves like a pressure-less dust ($\propto a^{-3}$) if it is averaged 
over a timescale sufficiently longer than $m^{-1}$ and sufficiently
shorter than $H^{-1}$.

For $J_2=0$, $\varphi$ behaves as a massive scalar field with mass $m$
and at low energy $H\ll m$, its energy density behaves like a
pressure-less dust ($\propto a^{-3}$) if it is averaged over a timescale
sufficiently longer than $m^{-1}$ and sufficiently shorter than
$H^{-1}$.

In summary we have shown that the stress energy tensor due to the motion
of a $\bar{D}3$-brane near the bottom of a warped throat region behaves
like dark matter $\rho\propto a^{-3}$. It is of course possible to 
consider motion of more than one $\bar{D}3$-branes as multi-component
dark matter.

\section{Dark matter production}
\label{sec:production}

Now let us discuss possible scenarios to generate the motion of
$\bar{D}$-branes as dark matter.

(i) {\bf $D/\bar{D}$ interaction at late stage of brane inflation:} 
In Ref.~\cite{KKLMMT} the attractive force between a $D$-brane and a
$\bar{D}$-brane were considered as the origin of the inflaton
potential. $\bar{D}$-branes are placed at the bottom of a throat and a 
mobile $D$-brane falls toward the bottom. This process inevitably
generates motion of not only the mobile $D$-brane but also all
$\bar{D}$-branes near the bottom of the throat since the $D/\bar{D}$
interaction acts between the mobile $D$-brane and each
$\bar{D}$-brane. If the number of $\bar{D}$-branes is more than one then
there remain remnant $\bar{D}$-branes after the brane inflation
followed by the annihilation of one $D/\bar{D}$ pair and reheating.

(ii) {\bf Closed string modes in the bulk:} 
The motion of $\bar{D}$-branes may be generated at reheating after
inflation if physics in the bulk of extra dimensions plays important
roles in the reheating process. For example, in Ref.~\cite{BBC} it was
suggested that in the context of the $D/\bar{D}$ brane inflation,
reheating on our brane placed at the bottom of a throat can be due to
closed string modes in the bulk produced during tachyon condensation. In
this case the closed string modes in the bulk can kick $\bar{D}$-branes
at the bottom and generate their motion.

(iii) {\bf Gravitational production:} 
It may also be possible to generate the dark matter within the context
of a $4$-dimensional effective theory without thinking about extra
dimensions. Gravitational production had been considered as a mechanism 
to produce superheavy dark matter (often called
WIMPZILLA)~\cite{Wimpzillas}. The same mechanism should work for the
production of the $\bar{D}$ dark matter if a $4$-dimensional effective
theory is valid at the time of production.

In all cases, it is expected that the total angular momentum along the
$S^2$ of the Klebanov-Strassler geometry should decay to an extremely
small value already during inflation, but if there are more than one
$\bar{D}$-branes then each $\bar{D}$-brane can obtain angular momentum
via the above mechanisms. (Of course the total angular momentum remains
extremely small.) In this case $\bar{D}$-branes may collide at late time
and lose their angular momenta since the sum of their angular momenta
are essentially zero. This may lead to intriguing cosmological and 
astrophysical consequences such as heavy particle production followed by
lepto- and/or baryo-genesis, generation of ultra high energy cosmic
rays, impact on the nucleo synthesis, etc. Further studies along this
line are worthwhile.

\section{Summary and discussion}
\label{sec:summary}

We have pointed out that in the warped string compactification, motion
of $\bar{D}$-branes near the bottom of a throat behaves like dark
matter. The existence of this type of dark matter is a necessary
consequence of the KKLT setup, where $\bar{D}$-branes are required to
uplift an AdS spacetime to a de Sitter spacetime.

Our arguments so far have been based on the implicit assumption that our
brane is somewhere in the compact manifold where the warp factor is of
order unity. This is the reason why we have written down the effective
action for the $\bar{D}$-brane in terms of the $4$-dimensional metric 
$g^{(4)}_{\mu\nu}$. Indeed, the physical (or induced) metric on our
brane in this setup is $g^{(4)}_{\mu\nu}$ up to an overall constant
factor of order unity. The dark matter mass $m$ is given in
(\ref{eqn:m2}) and can be estimated as
%
\begin{equation}
 \frac{m^2}{M_{10}^2} = \frac{2(4\pi)^{7/4}}{3I^{3/2}(0)}
  \cdot\frac{h^{-1/2}(0)}{g_s^{1/2}M}
  \simeq 3 \times h^{-1/2}(0)\cdot\left(\frac{g_s}{0.1}\right)^{1/2}
  \cdot\left(\frac{g_sM}{10}\right)^{-1},
\end{equation}
where we have used the formula $M_{10}^8=2/[(2\pi)^7{\alpha'}^4g_s^2]$
for the $10D$ gravity scale $M_{10}$. Hence, if we suppose the TeV
gravity, $M_{10}\sim TeV$, then the dark matter mass is lighter than TeV
by the factor $h^{-1/4}(0)$. Note that the warp factor $h^{-1/4}(0)$ is
given by
%
\begin{equation}
 h^{-1/4}(0) \sim \exp\left(-\frac{2\pi K}{3g_sM}\right)
\end{equation}
and can be exponentially small, where the positive integer $K$ is the
value of NS-NS flux required for moduli stabilization~\cite{GKP}.

On the other hand, if our brane is located near the bottom of the throat
then the dark matter mass should be different. The field $\varphi$ is no
longer canonically normalized with respect to the physical (or induced)
metric $\bar{g}^{(4)}_{\mu\nu}=h^{-1/2}(0)g^{(4)}_{\mu\nu}$, and the 
canonically normalized field is 
$\bar{\varphi}\equiv h^{1/4}(0)\varphi$. Correspondingly, the physical
mass of the field $\bar{\varphi}$ is $\bar{m}=h^{1/4}(0)m$ and, thus, 
%
\begin{equation}
 \frac{\bar{m}^2}{M_{10}^2} 
  \simeq 3 \times \left(\frac{g_s}{0.1}\right)^{1/2}
  \cdot\left(\frac{g_sM}{10}\right)^{-1}.
\end{equation}
Therefore, in this case the dark matter mass is around $TeV$ if we adopt
the TeV gravity.

We have suggested several scenarios for the dark matter production. One
of them is based on the $D/\bar{D}$ interaction at the late stage of
$D/\bar{D}$ inflation. Suppose that inflation in our $4$-dimensional 
universe is driven by the modulus representing the position of a mobile
$D$-brane relative to $\bar{D}$-branes near the bottom of a throat. As
argued in ref.~\cite{KKLMMT}, a successful inflation in this setup is
possible although fine-tuning is required to ensure the flatness of the
inflaton potential. The rolling of the inflaton is due to an attractive
force between the branes, and the same force inevitably generates the
motion of $\bar{D}$-branes as well. Since the $\bar{D}$-brane potential
has the minimum at the bottom of the throat, the $\bar{D}$-branes should
start moving around the bottom. When the inter-brane distance becomes
sufficiently short, the inflation ends and a pair of $D$- and
$\bar{D}$-branes annihilates. The $4$-dimensional universe should be
reheated by the energy released in the annihilation process. After the
annihilation of a $D/\bar{D}$ pair and the reheating of the universe,
the remaining $\bar{D}$-branes are still moving around the bottom. As we
have shown in this paper, the $\bar{D}$-brane motion around the bottom
behaves like dark matter in the $4$-dimensional universe. In this
context, the flatness of the inflaton potential requires that our brane
should be located somewhere in the Calabi-Yau manifold where the warp
factor $h^{-1/2}(\tau)$ is of order unity~\footnote{
A different possibility, where inflation can occur on our brane near
the bottom of a throat, will be discussed
elsewhere~\cite{Mukohyama-Kofman}.}. 
This means that the dark matter mass should be much lighter than the
$10D$ gravity scale, say $TeV$, as discussed above.

In order for this scenario to work, it must be made sure that the amount
of produced dark matter is not too much. Otherwise, the energy density
of the dark matter ($\propto a^{-3}$) would start dominating the 
radiation energy density ($\propto a^{-4}$) too soon. In the above
example, the production is due to the $D/\bar{D}$ interaction and the
interaction is very weak by definition: the enough inflation requires a
flat potential and the flat potential implies the weak interaction
between branes. From this consideration, it is expected that the amount 
of dark matter (i.e. the amplitude of $\bar{D}$-brane motion) after 
inflation should be very small. Very importantly, this expectation seems
compatible with the fine-tuning required for successful inflation since
both the smallness of the production rate and the flatness of the
inflaton potential are related to the weakness of the inter-brane
interaction. It is certainly worthwhile quantifying how much fine-tuning
this requires.

\section*{Acknowledgements}

The author would like to thank Y.~Imamura, L.~Randall and J.~Yokoyama
for useful discussions and comments. He would be grateful to
W.~Israel, H.~Kodama and K.~Sato for their continuing encouragement.

\end{document}